\documentclass[referee]{aa}
\usepackage{graphicx}
\begin{document}
\title{Radio asymmetry in 3C99}
%
% \thanks{We like to thank ... }
%
\author{F.~Mantovani\inst{1} 
 \and R.~Fanti\inst{1,2}
 \and C.~Ferrari\inst{1,3} 
 \and W.~Cotton\inst{4} 
 \and T.W.B. Muxlow\inst{5}
}
\institute{Istituto di Radioastronomia, CNR, Via P. Gobetti 101, Bologna, 
Italy 
\and Dipartimento di Fisica, Universit\'a degli Studi di Bologna, Bologna 
\and CERGA, CNRS, Observatoire de la Cote d'Azur, Nice, France 
\and National Radio Astronomy Observatory, Charlottesville, VA, USA 
\and Jodrell Bank Observatory, University of Manchester, UK
}
\offprints{Franco Mantovani,
\email{fmantovani@ira.cnr.it}
}
\abstract{ 
The N-galaxy 3C99, a Compact Steep-spectrum Source showing a triple
asymmetric structure, has been observed with several arrays of radio
telescopes at sub-arcsecond resolution. New images from MERLIN,
European VLBI Network and the Very Long Baseline Array show that 
the source components detected in the central region of 3C99 have
steep spectral indices, making it difficult to determine which of them
is the core. The asymmetric radio
structure of 3C99 is explained using the Scheuer-Baldwin 
continuous streaming model with a rather large difference in the
interstellar medium density on the two sides of the central region.
The age of 3C99 has been estimated; the source is found younger than
$10^6$ years. 
}
\date{Received / Accepted }
\maketitle
\section{Introduction}
\subsection{Radio properties of 3C\,99}
The radio source 3C99 (0358$+$004) is a powerful, 
$\sim10^{28}$Watt/Hertz at 408\,MHz N-galaxy at a redshift 0.426. 
On the arcsecond scale, imaged with 
the VLA and MERLIN,  it shows a triple asymmetric structure  
(Mantovani et al. 1990, 1997). The outer components are 
located rather asymmetrically relative to the nuclear region and show quite 
different luminosities and surface brightness. Along the major axis, 3C99 has 
an angular size of $\simeq$6 arcsec and a linear size of $\simeq$21 kpc 
($H_\circ=100$ Km sec$^{-1}$ Mpc$^{-1}$; q$_\circ=0$) almost fitting the 
selection criteria to be classified as Compact Steep-spectrum Source 
(CSS; see Fanti et al. 1995). 
The VLA images (Mantovani et al. 1997) show that  3C\,99 has a
rather asymmetric structure, with 4.9\% and 9.5\% 
of polarized emission at 5\,GHz and 8.4\,GHz respectively. 
The polarized emission comes mainly from
the bright hot spot $\sim$0.8 arcsec North-East to the
central component. The jet region (i.e. the region between these two
components) is also slightly polarized, with the electric vector 
perpendicular to the jet axis. In the hot spot region, a depolarization ratio
of 0.52 between 8.4\,GHz and 5\,GHz is seen. 
The position angle of the electric vectors is the same to within
the errors at the two frequencies, although there are $n\pi$ ambiguities.

The source 3C\,99 was observed with the European VLBI Network plus Green 
Bank and Haystack in 1984 at $\lambda$ 18cm (Mantovani et al. 1990), 
with an angular resolution of 11$\times$3 mas along a PA of 169$^\circ$. 
The central component was resolved into several components
of emission and 
shows a distorted structure. A compact feature was also 
detected in the hot spot, at $\sim$0.8 arcsec from the central
component. 

\subsection{IR, Optical and X-ray properties of 3C\,99}

The N-galaxy associated with 3C\,99 (Spinrad et al. 
1985) lies close to the central radio component labelled 
$C$ in Mantovani et al. (1990), which is expected to contain the nucleus 
of the source.
Observations with the IDS spectrograph of the Isaac Newton Telescope
with the slit of the spectrograph aligned along the radio
source major axis, have shown strong [OIII] emission and weak H$_\beta$.
The [OIII] emission extends 1 arcsecond 
above and below the nuclear region, with a higher redshift in the South-West 
direction, suggesting that thermal plasma in the North-East region is moving 
towards the observer, with $\Delta z \simeq 0.0025$ (Mantovani et al. 1990). 
3C\,99 has been observed again in 1993 with the ESO 2.2 metre telescope. 
The spectrum shows only narrow H$_\alpha$ and H$_\beta$ 
lines.  While most of the N-galaxies are actually classified BLRG, 3C\,99
is classified as a NLRG (Hes et al. 1996). 
Of particular interest is the superposition of the [OII]$\lambda$3727
distribution,  from which the continuum has been subtracted
(Hes, PhD Thesis, University of Groningen, 1996), on the VLA
A-array image at 4.9 GHz (Mantovani et al., 1990).
The [OII] nebular emission is about 7 arcsec in extent,
roughly centered on the 
nuclear radio region. The image shows 
that the north--east radio components are completely embedded in
the gas, while the south--west components are mostly outside  the
emitting gas (see Fig.\,1). 
Hes et al. (1996) also suggest that
the emission lines are due to photoionization by the central source and
that, if the B-model is adopted for photoionization, it is possible
to infer from the H$_\beta$ luminosity an electron density 
n$_e \leq 4\times10^2$ cm$^{-3}$ with a filling factor $\simeq 10^{-6}$. 
%
% ----------> mappa tesi Hes [OII] e VLA 6cm sovrapposte
%
\begin{figure*}
\centering
\includegraphics[width=12cm]{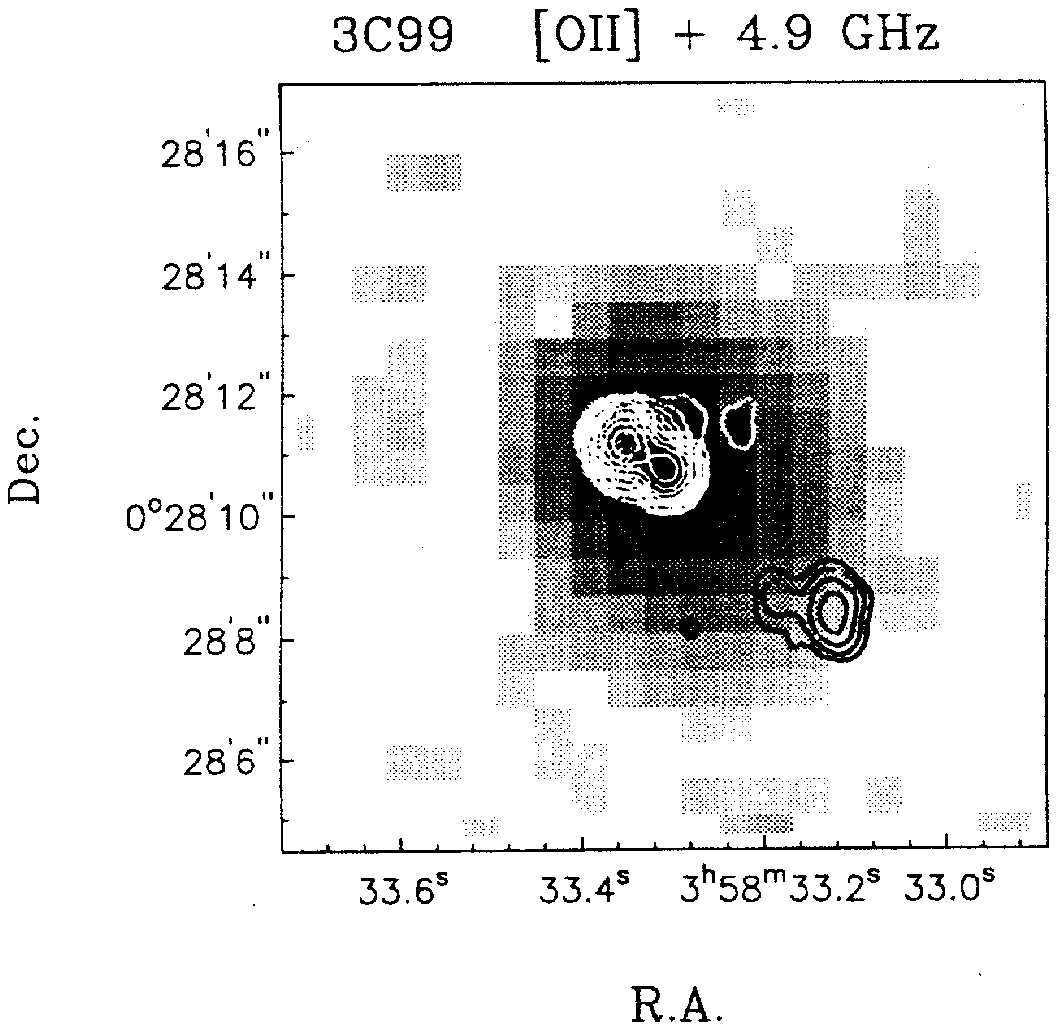}
\caption{The image of 3C\,99 (contours) at 4.9\,GHz (Mantovani et al. 1990)
super imposed to the [OII]$\lambda$3727 distribution (gray scale).
Contours are at 0.3 $\times$ (
--1, 1, 2, 4, 8, 16, 32, 64, 128, 256, 512) mJy/beam. The peak flux
density is 260 mJy. The beam is 0.48 $\times$ 0.42 arcsec$^2$ at 
PA 20$^\circ$. The image is reproduced from Hes, PHD Thesis, 1996). 
} 
\end{figure*}
N-galaxies are occasionally much brighter at 25 $\mu$m 
(van Bemmel \& Barthel 2001). The high IR luminosity would classify 3C\,99
as an unusually FIR-bright AGN. In addition certain CSS radio galaxies 
tend to be unusually FIR luminous (Hes et al. 1995), although this has not 
been found for the population as a whole (Fanti et al. 2000).  
On the other hand the X-ray luminosity is rather faint;
it was not detected by ROSAT (L$_X < 10^{43}$ ergs$^{-1}$; Crawford
and Fabian 1996). Together with the absence of broad lines, this
seems to exclude a quasar classification.
Table\,1 summarizes the observed properties of 3C\,99 at the various
wavelengths. 
\begin{table}
\centering
  \caption{\bf Summary of the observed properties at the various 
           $\lambda\lambda$ for 3C\,99}
\vspace{0.5cm}
\begin{tabular}{ll}
\hline
Optical Class                 & N-galaxy   \\
Red-shift                     & 0.426      \\
Radio Class                   & CSS        \\
Angular size                  & 6 arcsec   \\
Linear size                   & 21 kpc     \\
Polarization \% S$_{5GHz}$    & 4.9 \\
Polarization \% S$_{8.4GHz}$  & 9.5 \\
Radio Luminosity (408\,MHz)   & 10$^{28}$ Watts/Hz \\
IR Luminosity (60 $\mu$m)     & 10$^{45.4}$ ergs/sec \\
X Luminosity                  & $<$10$^{43}$ ergs/sec \\ 
\hline
\end{tabular}
\vspace{0.5cm}
\end{table}
\subsection{The present paper}
It has been suggested that the strong IR emission found in N-galaxies 
is mainly due to the interstellar medium (Hes et al. 1995), 
which in the case of 
CSSs sources might also be responsible for the radio plasma confinement. 
The asymmetry cannot be due to relativistic effects only, because the
closer lobe is also the brighter one, and the enhanced emission
in the northern lobe is also probably caused by a denser interstellar
medium.
We have carefully investigated the structure of 3C\,99 with high
resolution 
interferometric polarization observations.
Polarization observations could be a sensitive probe of the effects of 
compression or shocks on the structure of the magnetic field.

In the following we will present
observations done with the MERLIN and the European VLBI Network 
\footnote {The European VLBI Network is a joint facility of European and
Chinese radio astronomy institutes funded by their national research Councils}
at 5\,GHz, and with the VLBA 
\footnote {The Very Long Baseline Array and the Very Large Array are
facilities of the National Radio Astronomy Observatory, USA, operated
by Associated Universities Inc., under cooperative agreement with the
National Science Foundation.}  
at 1.6\,GHz and at 5\,GHz.

\section{Observations and data analysis}

The observations of 3C\,99 are summarized in Table\,2. 
The content of Table\,2 is as follows: 
column 1, array name; column 2, observing stations; column 3,
observing date; column 4, observing frequency; columns 5, 6 and 7, beam major
axis, minor axis and position angle respectively; column 8, total intensity
rms noise; column 9, polarized intensity rms noise.
\begin{table}[h]
\centering
  \caption[ ]{\bf Observational parameters of 3C\,99}
\vspace{0.5cm}
\begin{tabular}{llllccccc}
\hline
Array   & Stations$^a$        & Date  & Freq.&     &  Beam  &        &$\sigma_t$& $\sigma_p$   \\
        &                     &       & MHz  & mas &  mas   & deg & mJy/beam & mJy/beam  \\
\hline
EVN     &Ef,Mc,Nt,On-60,Jb-2,Wb    &1990.87& 4975 & 7.1 &  4.4   &  65    & 0.15     &               \\
EVN     &Ef,Mc,Nt,On-60,Jb-2,Wb,Sm &1995.38& 4975 & 8.4 &  4.5   &  37    & 0.16     &             \\
MERLIN  &All                  &1995.38& 4994 &  62 &  54    &  33    & 0.3      &            \\
MERLIN-Cm&All--Cm       &1995.38& 4994 &100 & 100    &        & 0.5      &   0.5       \\
VLBA    &All--Sc+VLA1   &1998.74& 1655 &17.5 & 4.9    & --21   & 0.9      &   0.2       \\
VLBA    &All--Sc+VLA1   &1998.74& 4619 & 5.6 &  2.0   & --18   & 0.2      &   0.3        \\
\hline
\multicolumn{9}{l}{ } \\
\multicolumn{9}{l}{
$^a$: The station labels listed are as follows: Ef = Effelsberg (D) 100-m, 
Mc = Medicina (I) 32-m,  } \\ 
\multicolumn{9}{l}{
Nt = Noto (I) 32-m, On-60 = Onsala (S) 20-m, Jb-2 = Jodrell Bank (UK) 25-m, 
Cm = Cambridge (UK) 32-m, } \\ 
\multicolumn{9}{l}{ 
Wb = Westerbork array (NL) n$\times$25-m, Simeiz (Ukraine) 22-m; 
VLBA: Sc = St. Croix.}
%      Hn = Hancock,
%      Fd = Fort Davis, Kp = Kitt Peak, La = Los Alamos, Nl = North Liberty,
%      Pt = Pie Town, Br = Brewster, Ov = Owens Valley, (USA) 25-m.
%
\end{tabular}
\end{table}
\subsection{ MERLIN observations }
In 1995.38 the source was observed by the EVN+MERLIN array.
In order to calibrate the  MERLIN data set, initial values for the telescope
and correlator gains were determined from a short observation of a bright,
unresolved calibration source. The primary calibrator 3C286 was
included during the observations in order to fix the position angle of 
the electric vector. The absolute flux density was scaled assuming a flux 
density of 2325.6\,mJy for OQ208. The source DA193 was used to calculate
the leakage term.
Images from the MERLIN LHC total intensity and polarization data, the
latter
without the Cambridge data due to the limits in bandwidth of the radio link, 
were produced using ${\cal AIPS}$ and shown in Figs 2 and 3 respectively.
%

% ------------->   2 merlin maps : 1 full resolution; 1 without Cm with pol
\begin{figure*}
\centering
\includegraphics[width=12cm]{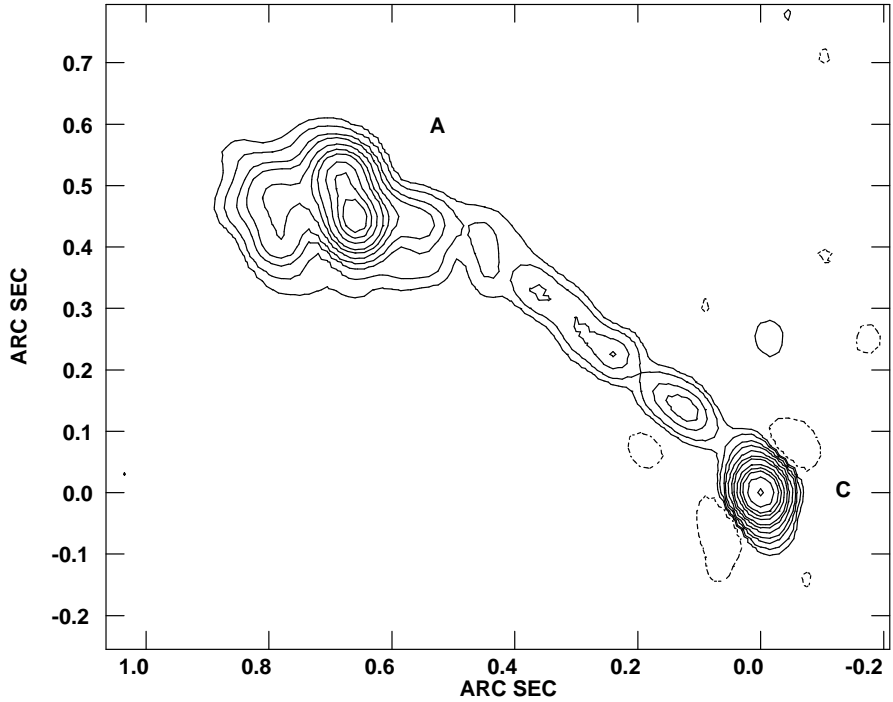}
\caption{The MERLIN (epoch 1995.38) image of 3C99 at 5\,GHz. Contours are at
--1, 1, 2, 4, 6, 10, 15, 20, 30, 40, 50, 70, 100 mJy/beam. The peak flux
density is 105.6 mJy. The beam is $62 \times 54$ mas at PA 33$^\circ$. 
} 
\end{figure*}
\begin{figure*}
\centering
\includegraphics[width=12cm]{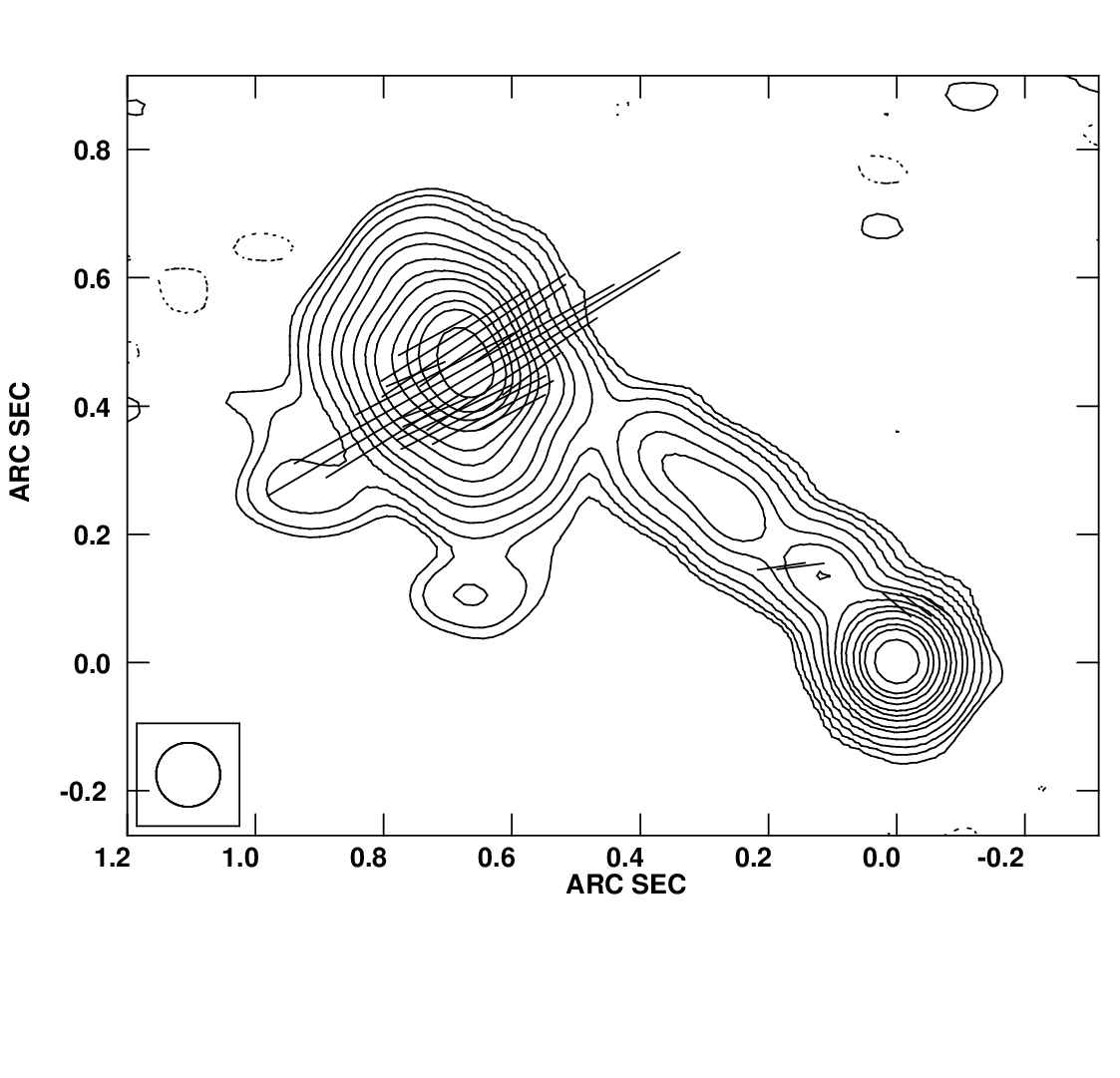}
\caption{The MERLIN minus Cambridge (epoch 1995.38) image of 3C99 at 5\,GHz. 
Contours are at
--0.25, 0.25, 0.5, 1, 2, 4, 6, 10, 15, 20, 30, 40, 50, 70, 100) mJy/beam. 
The peak flux
density is 96.4 mJy. The beam is $100 \times 100$ mas. A electric vectors
length $1\arcsec=20$ mJy/beam.
} 
\end{figure*}
The images in Fig.s\,2,3 show that the core region (component $C$) and the 
North-East hot-spot (component $A$) are connected by a long, straight, 
collimated jet. The jet bends and slightly increases in transverse size
before reaching the hot-spot region, which is resolved
into a triple hot-spot. Component $C$ is point-like at this resolution and
is not polarized. The first knot along 
the jet is in fact polarized $\sim8$\%, while in the remaining part of
the jet, the polarization, if any, is below the detection limit. 
The region $A$ shows $\sim6.4$\% polarized flux density in total.
% Component $C$ is weakly polarized, $\sim1.7$\%, with the
% polarized flux coming from the outer part of the component, nearby the
% place where the jet begins.
In the south--west region, the weak and extended emission seen with the VLA 
observations (Mantovani et al., 1990, 1997 and Fig.\,1), was not detected 
by MERLIN.

Mantovani et al. (1997) using VLA observations measured a strong 
depolarization (0.52) between 8.4\,GHz and 5\,GHz and no indication of
Faraday Rotation of the electric vector. Comparing the MERLIN image
with the VLA images a similar percentage of polarized flux density is
found at 5\,GHz for the component A, confirming the depolarization between 
8.4\,GHz and 5\,GHz. In the MERLIN image,
the mean electric vector position angles around the peak of polarized
emission are $120\pm1$ deg in agreement with previous results.
\subsection{ VLBA observations }
The source was observed with 9 VLBA antennas (all but Saint Croix)
and 1 VLA antenna for about 10 hours, switching 
between 1.6\,GHz and 5\,GHz. We made use of four 8\,MHz bands at each
frequencies centered at  1651.49, 1659.49, 1667.49 and
1675.49; and 4615.49, 4653.49, 4850.49, 5090.49 respectively. 
The target source was observed together with the calibrator
sources BL\,Lac, 3C84, 3C138 and OJ287. The data have been analyzed
using ${\cal AIPS}$.
The data have been calibrated in amplitude and phase, the LHC-RHC delay 
difference and the instrumental polarization have been corrected by
fringe fitting a segment of the cross-hand data from a strong calibrator
(3C\,84).  

The two data sets were then averaged in frequency over the four IF
channels for a 
total observing bandwidth of 32\,MHz. The images obtained at 1.6\,GHz are
shown in Fig.\,4 and Fig.\,5. The latter figure is a closeup of
the central region of 3C99.
The hot-spot region has been resolved into  three 
components which appear resolved and off--axis by $\sim15^\circ$ 
with respect to the central region, as also apparent in the MERLIN map.

The central region shows two bright and two weak outer components. 
At 5\,GHz we detected the central component only, which is dominated by 
the two bright components seen at 1.6\,GHz. 
A similar image is seen at 4.6\,GHz  in Fig.\,6 where they appear extended
due to the better resolution. Again the structure is dominated by two
bright blobs. It is worth noting that
none of the components of Fig.\,5 is point--like. 

No polarized emission was detected, at the two frequencies, 1.6\,GHz 
and 5\,GHz above the rms noise of 0.2 mJy/beam and 0.3 mJy/beam respectively.
%
%
%
% ------------------>  2 mappe vlba 18cm
\begin{figure*}
\centering
\includegraphics[width=12cm]{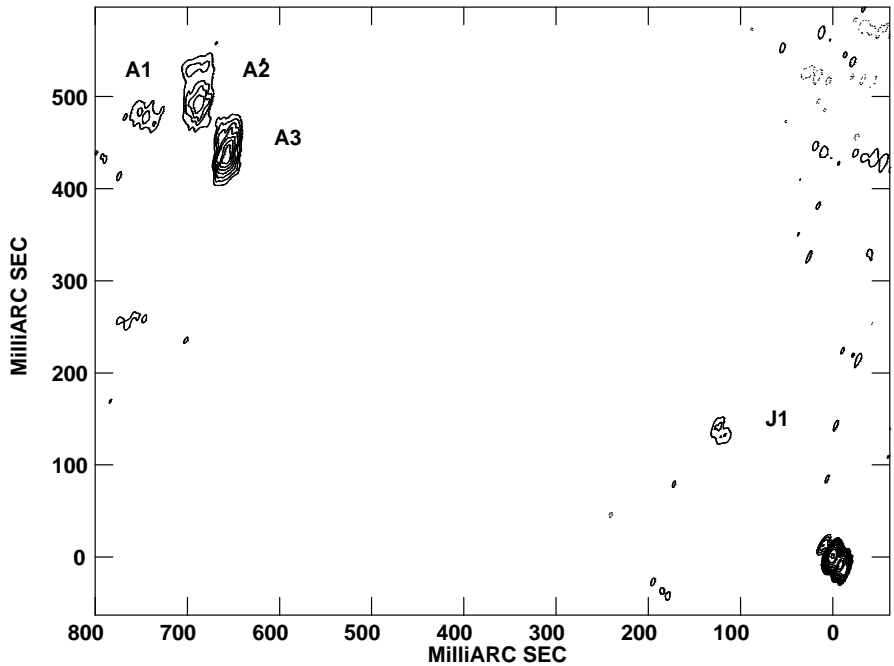}
\caption{The VLBA (epoch 1998.74) image of 3C99 at 1.6\,GHz. Contours are at
--2, 2, 3, 4, 5, 6, 7, 8, 10, 15, 20, 30, 40, 60 mJy/beam. The peak flux
density is 63.6 mJy. The beam is 17.5 $\times$ 4.9 mas at PA --21$^\circ$.
} 
\end{figure*}
\begin{figure*}
\centering
\includegraphics[width=12cm]{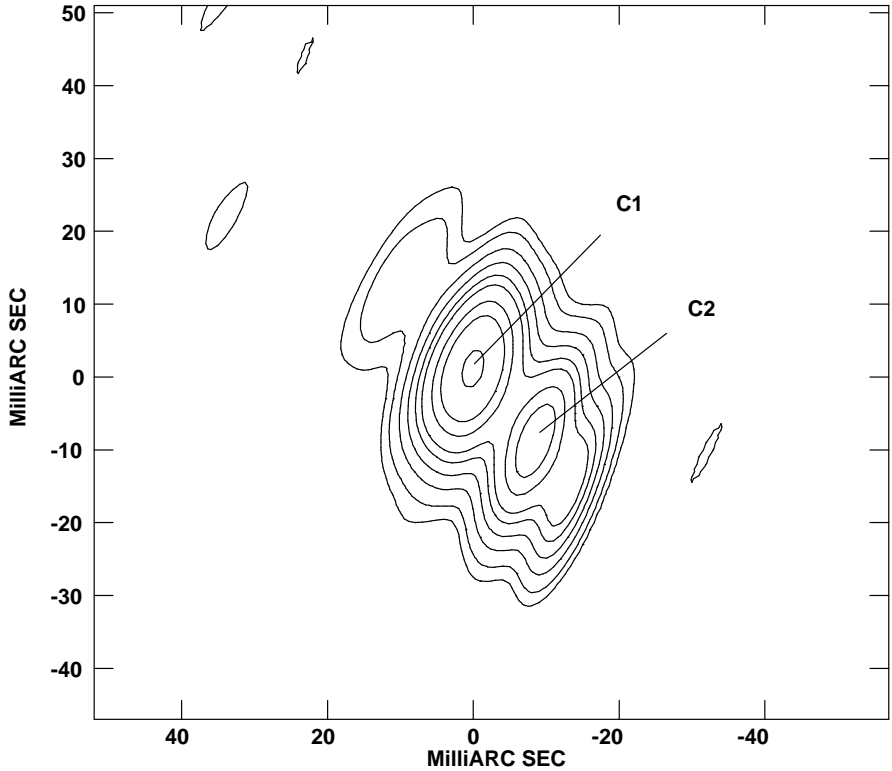}
\caption{The VLBA (epoch 1998.74) image of the central region of 3C99 at 
1.6\,GHz. Contours are at --1.5, 1.5, 3, 6, 10, 15, 20, 30, 40, 60 mJy/beam. 
The peak flux density is 63.6 mJy. The beam is 17.5 $\times$ 4.9 mas 
at PA --21$^\circ$.
} 
\end{figure*}

% ------------------>  1 mappa vlba 6cm
\begin{figure*}
\centering
\includegraphics[width=12cm]{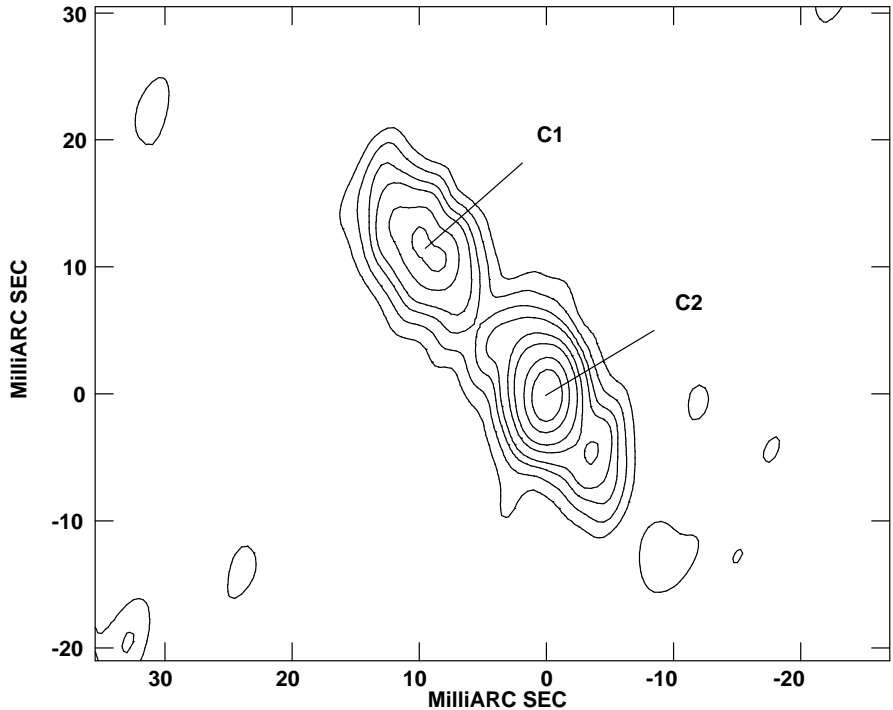}
\caption{The VLBA (epoch 1998.74) image of 3C99 at 5\,GHz. Contours are at
--0.6, 0.6, 1, 1.5, 2, 3, 4, 6, 8 mJy/beam. The peak flux
density is 10.5 mJy. The beam is 5.6 $\times$ 2.0 mas at PA --18$^\circ$.
} 
\end{figure*}
\subsection{ EVN observations }
The two EVN observations at 5\,GHz were done 
recording in MarkIII  Mode  A (56\,MHz bandwidth; date 1990.87) and
recording in MarkIII  Mode B (28\,MHz bandwidth; date 1995.38).
The data were processed at the Max-Planck-Institut f\"ur Radioastronomie 
Correlator in Bonn. The correlator output was calibrated in amplitude and 
phase using ${\cal AIPS}$ 
\footnote {${\cal AIPS}$ is the NRAO's {\it Astronomical Image Processing 
System}} and imaged using DIFMAP \footnote {DIFMAP is part of the 
{\it Caltech VLBI software Package}} (Shepherd et al.\ 1995).
Total power measurements and gain-curve of the telescopes, were applied in 
the amplitude calibration process.
The self-calibration procedure, which uses closure amplitudes to determine
telescope amplitude corrections, gave calibration factors that were within
10\% of unity for all the telescopes.
%
% ---------->  2 mappe evn 6 cm

\begin{figure*}
\centering
\includegraphics[width=12cm]{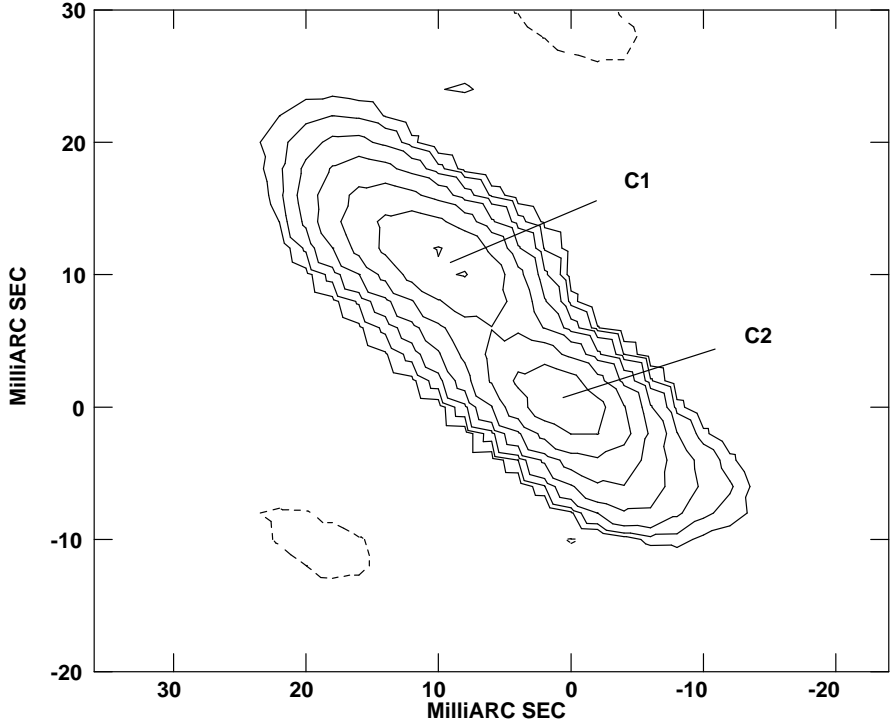}
\caption{The EVN (epoch 1990.87) image of 3C99 at 5\,GHz. Contours are at
--0.4, 0.4, 0.8, 1.6, 3.2, 6.4, 12.8, 25.6, 51.2 mJy/beam. The peak flux
density is 43.3 mJy. The beam is 7.1 $\times$ 4.4 mas at PA 65$^\circ$.
} 
\end{figure*}
\begin{figure*}
\centering
\includegraphics[width=12cm]{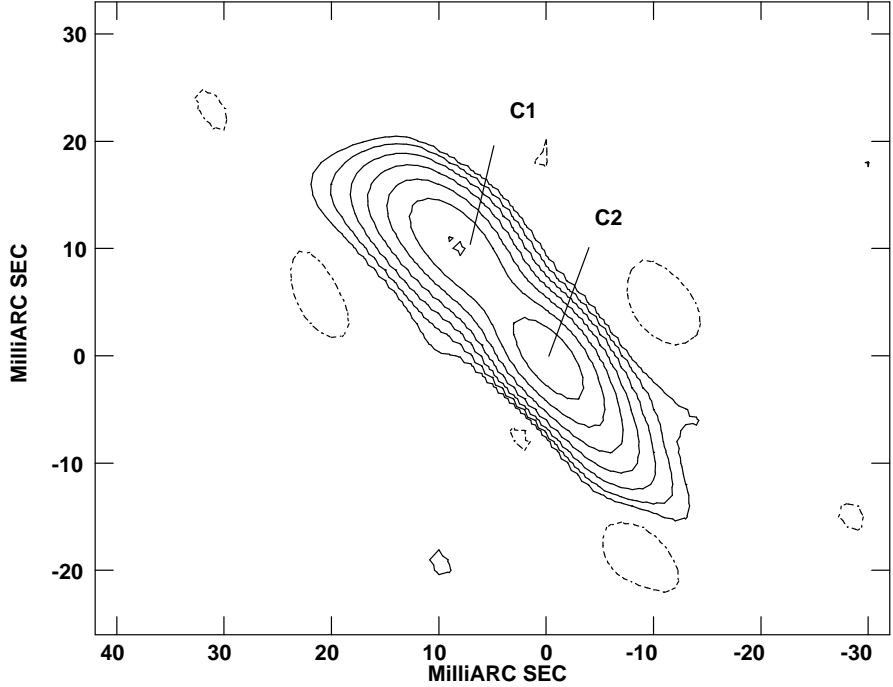}
\caption{The EVN (epoch 1995.38) image of 3C99 at 5\,GHz. Contours are at
--0.4, 0.4, 0.8, 1.6, 3.2, 6.4, 12.8, 25.6, 51.2 mJy/beam. The peak flux
density is 43.7 mJy. The beam is 8.0 $\times$ 4.5 mas at PA 62$^\circ$. 
} 
\end{figure*}
The two images from the two observing sessions 
are shown in Fig.\,7 and Fig.\,8 respectively. They look very similar
in structure, size and flux density despite  almost five years 
between the two observations. We see two resolved, elongated 
components, aligned with the source major axis in PA$\sim 40^\circ$, of almost 
equal flux density.  The properties of 3C\,99, from the 
EVN and the VLBA observations are given in Table\,3.
\begin{table}
\centering
  \caption{\bf Observed properties of the compact components of 3C\,99}
\vspace{0.5cm}
\begin{tabular}{lrrrrrcrrrrr}
\hline
Array& Date  &Freq.&    & Beam&    &Comp.&$S_{peak}$&$S_{tot}$&Major&Minor&PA\\
     &       &MHz  & mas&  mas& deg &     & mJy/b   & mJy   & mas & mas &deg\\
\hline
EVN  &1990.87&4975 & 7.1&  4.4&  65 &$C1$ & 27.2     & 44.4  & 7.5 & 1.4 &52 \\
     &       &     &    &     &     &$C2$ & 43.6     & 53.1  & 4.9 & 0.7 &46 \\
EVN  &1995.38&4975 & 8.4&  4.5&  37 &$C1$ & 26.2     & 46.6  &10.1 & 2.4 &44 \\
     &       &     &    &     &     &$C2$ & 43.1     & 55.9  & --  & --  &-- \\
% MERLIN  &1995.38& 4994&  62 &  54   &  33 & 0.3      &            \\
% MERLIN-CB&1995.38&4994&100 & 100    &     & 0.5      &   0.5       \\
VLBA &1998.74&1655 &17.5& 4.9 &--21 &$A1$ & 3.8      & 20.4  &49.1 &24.0 & 54\\
     &       &     &    &     &     &$A2$ & 4.6      & 84.3  &95.5 &29.7 &178\\
     &       &     &    &     &     &$A3$ & 8.5      & 98.8  &58.5 &23.0 &169\\
     &       &     &    &     &     &$J1$ & 3.3      & 18.2  &49.1 &17.1 & 34\\
     &       &     &    &     &     &$C1$ & 62.0     &138.5  & 8.7 & 5.3 & 51\\
     &       &     &    &     &     &$C2$ & 44.7     &127.3  &12.9 & 6.6 & 28\\
VLBA &1998.74&4619 & 5.6&  2.0&--18 &$C1$ &  4.3     & 23.4  & 9.8 & 3.8 & 40\\
     &       &     &    &     &     &$C2$ & 10.0     & 26.7  & 5.3 & 2.4 & 32\\
     &       &     &17.5& 4.9 &--21 &$C1$ & 13.1     & 20.0  & --  & --  &-- \\
     &       &     &    &     &     &$C2$ & 22.4     & 34.9  & --  & --  &-- \\
\hline
\end{tabular}
\vspace{0.5cm}
\end{table}
\subsection{Comparison of the VLBI images of the central component}

The 5\,GHz image has been convolved with the 1.6\,GHz beam and
the flux densities of the two main components compared. A spectral index 
of $\alpha_1=0.8$ is derived for component $C1$ and
$\alpha_1=1.8$ for component $C2$ (S$\propto\nu^{-\alpha}$). 

To check for any possible change of the relative position of the two 
components $C1$ and $C2$ with epoch, we made use of the images at 5\,GHz 
(their maximum time separation is $\sim8$ years), all convolved to the 
same beam size.
The angular separation between the two components of $\sim14$ mas did 
not change within the estimated errors of $\simeq$ 0.5 mas. The errors 
are affected by the 
discrete size of the components (of the order of the beam size) which
are extended in the same direction as the source major axis. 

The derived parameters, resolved angular sizes, steep spectral indices, 
constant separation, suggest that nor $C1$ neither $C2$ can
be considered as the core of 3C99. 
\section{Discussion}

The arcsecond resolution images of 3C99 show that the source has a
triple asymmetric structure.
Even if we could not achieve a firm identification  of the
source core, the core location is doubtlessly in the component $C$
region. Also,
the position of the optical N-galaxy 3C99 (Spinrad et al. 1985) is close to
that component. Thus, we know approximately where the central engine is
located and we can confirm the source asymmetry.
The ratio between the distances
of the hot-spots from the center of component $C$ is 4.5, while that of 
the hot-spots brightnesses is $\approx$40.
From the optical observations we also know that the North-East side is 
approaching the observer being less red-shifted than the South-West part 
(see Mantovani et al. 1990). 

Asymmetry in the radio lobe arm length ratio is expected when the hot
spots move at a relativistic speed as a consequence of the difference in 
travel time of the radiation emitted by the approaching and receding hot
spot.
The approaching lobe should be the one seen at a later stage of evolution and 
therefore more distant from the core as compared to the receding 
one.
Furthermore some relativistic boosting (de-boosting) should also be 
expected for the approaching (receding) hot spot, giving rise as well
to an asymmetry in luminosity.
In 3C99 the lobe closer to the core is the brighter and approaching 
one. It is clear that relativistic effects alone cannot fully explain the 
observed asymmetry.

\smallskip

Alternatively, we attribute the asymmetry to an inhomogeneity in the ambient
medium on the two sides of the source. The brighter and closer lobe would be
the one moving through a more dense medium having a small density gradient, 
being slowed down and remaining bright as a consequence of a more effective
confinement.
An indication that this could
be the case comes from the image of Fig.\,1, a superposition of the
radio and [OII] images. If we assume that they actually overlap in space,
it follows that 
the north--east bright component of the source, i.e. the part closest
to the core, is completely immersed in the cloud of gas, while the 
south--west hot--spot lies in a 
region at the border where the gas density is clearly much lower. 

\subsection {A model for asymmetry.}

To explain the asymmetry in the radio structure of 3C99 we made use of 
the Scheuer-Baldwin model (Scheuer 1974; Baldwin 1982), one of the so called
continuous streaming model. In the model, a relativistic twin-jet propagates 
in an external medium with a non uniform density.

We assume that the jets carry an energy flux $F_e$ and a thrusts $F_e/c$.
The advance speed of the jets head (hot spots), $v_h$, is controlled by the 
balance between the jet thrust and the ram-pressure of the ambient medium, 
viz. $v_h \propto \frac{F_e}{c} D^{-2}$.
Modeling the external density as

$\rho \propto D^{-n}$

\noindent

where $D$ is the distance from the nucleus. By integration of these 
relations, we get the instantaneous and average advance speed 

$v_h \propto D^{-(2-n)/2}~~~~~$ $~~~<v_h> = v_h \times (4-n)/2$ 

The energy carried by the jet is assumed to  accumulate into the lobe 
under equipartition conditions. This allows the computation of the
lobe luminosity (Begelman 1996) as 

$L_l \propto [F_e \times \tau \times f(n)]^{7/4} \times V_l^{-3/4}$

where $\tau$ is the lobe age, {\it f(n)} a factor accounting for the energy 
lost in the source expansion (of the order of 0.5) and $V_l$ the lobe
volume.
%(f(n=2) = 0.5; f(n=0) = 0.63)

Finally, assuming again equipartition, we evaluate the hot spot luminosity as

$L_h \propto F_e^{7/4} \times r_h^{-0.5}$

\noindent
where $r_h$ is the hot spot radius, that we assume, for sake of simplicity,
proportional to $D$.

We remark that, because of the large advance speeds, the travel time effects 
would cause the two lobes to be seen at different ages, thus

$\tau_1/\tau_2 = (1+<\beta_2> cos \theta)/(1-<\beta_1> cos \theta)$

Therefore, an important item for the following discussion is the 
orientation of the source with respect to the line of sight.
\subsection{Physical parameters from jets, lobes and hot spots} 
In the above framework, we will interpret the asymmetry in 3C99 making use of
the physical quantities, listed in Table\,4, derived from the present
observations and from those presented in Mantovani et al. (1990). 
\begin{table}
\centering
  \caption{\bf Physical quantities from the images of 3C\,99}
\vspace{0.5cm}
\begin{tabular}{lcccccccc}
\hline
Region & Comp. & log$\mbox{\rm P}_{\rm 1.4~GHz}$&$\mbox{\rm AS}_{1}~{\times}~\mbox{\rm AS}_{2}$ & $\mbox{\rm LS}_{1}~{\times}~\mbox{\rm LS}_{2}$& Volume& $\mbox{\rm U}_{min}$& $\mbox{\rm u}_{min}$& $\mbox{\rm H}_{eq}$\\
\hline
&  &  & mas & kpc& ${\rm kpc}^3$& erg& ${\rm erg}{\rm cm}^{-3}$& mGauss\\
\hline
Lobe$_{North}$& & 26.4& 750$\times$470 & 4.1$\times$2.8& 17& 1.8$\times10^{57}$& 3.5$\times10^{-9}$& 0.2\\
\hline
Lobe$_{South}$& & 25.1& 1500$\times$1500& 5.9$\times$5.9& 111& 0.8$\times10^{57}$& 2.4$\times10^{-10}$& 0.05\\
\hline
Hot-spot$_{North}$& I& 25.5& 52$\times$42& 0.32$\times$0.25& 0.01& 2.2$\times10^{55}$& 7.0$\times10^{-8}$& 0.9\\
                  &II& 25.7& 48$\times$42& 0.29$\times$0.23& 0.01& 2.6$\times10^{55}$& 9.6$\times10^{-8}$& 1.0\\
                 &III& 25.2& 148$\times$87& 0.90$\times$0.50& 0.12& 4.2$\times10^{55}$& 1.1$\times10^{-8}$& 0.35\\
\hline
Hot-spot$_{South}$& I& 24.2& 531$\times$290& 3.15$\times$1.72& 4.9& 5.0$\times10^{55}$& 3.6$\times10^{-10}$& 0.06\\
                  &II& 23.9& 300$\times$231& 1.0$\times$0.76& 1.75& 2.2$\times10^{55}$& 4.4$\times10^{-10}$& 0.07\\
\hline
Jet (mas)         &  & 25.3& 700$\times$70& 4.14$\times$0.21& 0.09& 4.5$\times10^{55}$& 1.6$\times10^{-8}$& 0.4 \\
\hline
5\,GHz            &C1& 25.0& 10.4$\times$4.2& 0.05$\times$0.02& 1.8$\times10^{-5}$& 8.0$\times10^{53}$& 1.5$\times10^{-6}$& 4.1\\
1.6\,GHz          &C1& 25.4& 8.6$\times$3.1& 0.03$\times$0.01& 1.8$\times10^{-5}$& 1.4$\times10^{54}$&
 2.8$\times10^{-6}$& 5.5\\
\hline
5\,GHz            &C2& 25.3& 10.5$\times$3.6& 0.05$\times$0.02& 0.9$\times10^{-5}$& 8.6$\times10^{53}$& 3.3$\times10^{-6}$& 6.0\\ 
1.6\,GHz          &C2& 25.3& 29.4$\times$4.3& 0.18$\times$0.01& 5.0$\times10^{-5}$& 1.8$\times10^{54}$& 1.2$\times10^{-6}$& 3.6\\
\hline
\multicolumn{9}{l}{ } \\
\multicolumn{9}{l}{
The quantities have been derived assuming equipartition, a filling } \\
\multicolumn{9}{l}{
factor of 1, electron to proton ratio of 1 and the behavior of the radio } \\
\multicolumn{9}{l}{
spectrum between 10\,MHz and 100\,GHz. To determine the flux densities of } \\
\multicolumn{9}{l}{
the north--east lobe, we subtracted the core, jet } \\
\multicolumn{9}{l}{
and hot spot flux densities from images at higher resolutions (for example the MERLIN } \\
\multicolumn{9}{l}{
image in Fig.\,2). Since those images were also at different frequencies, } \\
\multicolumn{9}{l}{
the flux densities were scaled accordingly, adopting a mean spectral } \\
\multicolumn{9}{l}{
index 0.5 (S$\propto \nu^{-\alpha}$) } \\
\multicolumn{9}{l}{
The linear dimensions were derived from the angular dimensions obtained } \\
\multicolumn{9}{l}{
using the {\cal AIPS} task JMFIT, deconvolved from the beam.  Table\,4 is } \\
\multicolumn{9}{l}{
organized as follows: column 1, region; column 2, component (the hot-spot } \\
\multicolumn{9}{l}{
north components I, II and III correspond to the components A1, A2 and A3 } \\
\multicolumn{9}{l}{
of Fig. 4); column 3, radio power at 1.4\,GHz; column 4, linear dimensions } \\
\multicolumn{9}{l}{
in kpc; column 5, volume in kpc$^3$; column 6, minimum energy in erg; } \\
\multicolumn{9}{l}{
column 7, energy density erg/kpc$^3$; column 8, magnetic field in mGauss. } 
\end{tabular}
\end{table}

The ratio of the distances of the lobes at the two opposite sides of the nucleus 
is: 
\begin{equation}
\frac{D_1}{D_2} =  \frac{<\beta_1>}{<\beta_2>} \times 
                  \frac{1 + <\beta_{2h}> cos\theta}{1 - <\beta_{1h}> cos\theta}
\end{equation}
where $D_1$ and $D_2$ are the distances of the approaching and receding lobe
respectively, $<\beta_1>$ and $<\beta_2>$ their mean velocities relative to
the speed of light and $\theta$ the angle between the source major 
axis and the observer's line of sight. The second factor accounts for
differences in the travel time from the lobes.

The lobe luminosity ratio is given by:

\begin{equation}
\frac{L_{l1}}{L_{l2}} =  [\frac{F_{e1}}{F_{e2}} \times
    \frac{1 + <\beta_2> cos\theta}{1 - <\beta_1> cos\theta}  
    \frac {f(n_1)}{f(n_2)}]^\frac{7}{4}
    \times (\frac{V_{l2}}{V_{l1}})^\frac{3}{4}
\end{equation}
The factor $(1 + <\beta_2> cos\theta)/(1 - <\beta_1> cos\theta)$
accounts for the fact that the two lobes are seen at
different times of their evolution.
A similar equation holds for the hot spot luminosity ratio:
\begin{equation}
\frac{L_{hs1}}{L_{hs2}} =  (\frac{D_2}{D_1})^\frac{1}{2}
   (\frac{F_{e1}}{F_{e2}})^\frac{7}{4} 
   [\frac{\gamma_2 (1 + \beta_2 cos\theta)}
         {\gamma_1 (1 - \beta_1 cos\theta)}]^{3+\alpha_h}
\end{equation}
where $\gamma = (\sqrt{1-\beta^2})^{-1}$ is the Lorenz factor,
$\alpha_h$ the hot spot spectral index.
The parameters needed to satisfy the above equations
can be compared to those derived from the existing observations of 3C99, 
summarized in Tables\,4 and 5.
\begin{table}
\centering
  \caption{\bf Physical parameters of 3C\,99}
\vspace{0.5cm}
\begin{tabular}{lllll}
\hline
Parameter            & Image Freq. & Approaching jet & Receding jet & Ratio  \\
\hline
Distance nucleus-lobe& 1.4\,GHz & $D_1= 1260$mas   & $D_2= 5940$mas& 0.21   \\
Lobe Flux Density    & 0.4\,GHz & $S_{l1}= 2900$mJy& $S_{l2}= 170$mJy& 17.1 \\
Hot-spots Flux Density& 5\,GHz  & $S_{hs1}= 159$mJy& $S_{hs2}= 4$mJy & 37.9 \\
Jet Brightness  &     & $b_j\approx 4$mJy/beam&$b_{cj}\leq 0.32$mJy/beam& $>10$ \\
\hline
\multicolumn{5}{l}{ } \\
\multicolumn{5}{l}{
The jet counter-jet brightness ratio in the last row of Table\,5 has been } \\
\multicolumn{5}{l}{
derived from the MERLIN image at 5\,GHz considering the jet as having a } \\
\multicolumn{5}{l}{
conical structure with longitudinal dimension of $\sim$700\,mas, a transverse} \\
\multicolumn{5}{l}{
dimension of $\sim$70\,mas, an angle $\leq 1^\circ$ and a total flux } \\
\multicolumn{5}{l}{
density of 43\,mJy. The jet mean brightness $b_j\simeq 4$ mJy/beam is } \\  
\multicolumn{5}{l}{
obtained dividing the total flux density by the beam area to jet area } \\
\multicolumn{5}{l}{
ratio. Since the counter-jet is below the detection limit of the } \\
\multicolumn{5}{l}{
MERLIN image, we have $b_{cj} \leq 3\sigma = 0.32$ mJy/beam, and a } \\
\multicolumn{5}{l}{
ratio between the brightnesses $\geq 12$. }
\end{tabular}
\end{table}
\subsection{Search for the unknown parameters}
We  make use of the above equations to evaluate  the unknown
parameters $\beta_1$, $\beta_2$, $n_1$, $n_2$ and $\theta$.

We first constrain the range of values for $\theta$.
Assuming a symmetric two-sided jet, the lower limit to the jet to 
counter-jet brightness ratio implies $\beta_j cos\theta \simeq 0.63$, 
which requires $\theta \leq 50^\circ$.  Moreover, in order to
fit the unified models scheme (see Saikia et al. 1995) a CSS source optically
classified as a N-galaxy is expected to have the radio axis at
$30^\circ < \theta < 50^\circ$ to the observers line of sight.

In order to maximize the effects due to the asymmetric distribution of the
ambient medium, we assume $n_1$ = 0 and $n_2$ = 2. Following the models
of Scheuer (1974) modified by Baldwin (1982) and of Begelman (1996), we
compute $f(n_1)/f(n_2)$ $\approx$ 1.25.
Furthermore $\beta_2 = <\beta_2>$ and $\beta_1 = 0.5 <\beta_1>$.

The observed parameters which the model uses are:

$D_1$/$D_2$ = 0.21 $\pm 0.02$

$L_{l1}$/$L_{l2}$ = 23 $\pm 4$

$V_{l1}$/$V_{l2}$ = 25 $\pm8$

$L_{hs1}$/$L_{hs2}$ = 35 $\pm$ 10

$r_{h1}$/$r_{h2}$ = 8 $\pm 2$

One of the most uncertain input parameter is the ratio of the lobe volumes,
deduced from the 1.7 GHz map of Mantovani et al. (1990), which is the best
in order to evaluate the lobe's luminosity, but suffers from poor angular
resolution for a proper measurement of the lobe's size.
On the other hand, the above mentioned source models (Scheuer-Baldwin 
and Begelman) allow the
computation of the volume ratio, which is consistent with the observed one.

An additional uncertainty comes from the identification of the $hot spot$ 
of the north lobe. If we take the brighter one (named I in Tab.\,4, 
labelled A1 in Fig.\,4) we get a luminosity ratio of 30, while using
the second  
brightest (named II in Tab.\,4, labelled A2 in Fig.\,4) we get a ratio
of 20.
The previous equations allow a search over a range of parameters
to account for the observed asymmetry of the radio source.
For $30^\circ \leq \theta \leq 40^\circ$, the first two equations require
$ <\beta_2> \approx 0.3 \pm 0.1$ and $<\beta_2>/<\beta_{1}> \approx 5.5$.

The above numbers allow the computation of the ratio of hot spot luminosities,
with equation (3). We get: $10<L_{h1}/L_{h2} <15$.
This range of values is somewhat lower than the one we have from the
observations ($\approx 35$).
In order to increase the $L_{h1}/L_{h2}$ computed ratio, one should increase
the  $\beta$s, which in turns implies an increase of the ratio
$L_{l1}/L_{l2}$.
\subsection{The age of 3C99}
The radio source age can be derived from the age of the emitting
relativistic electrons providing we know the break frequency in the
spectral index of the source. From the
existing images, we derive straight steep spectral indices for the two lobes
($\alpha \sim 1.2$), thus
we cannot infer any value for break frequencies. The only suggestion we
can made is that $\nu_{break}\leq 408$\,MHz. For the equipartition
magnetic fields of Table\,4, following Carilli et al. (1991) and
Murgia et al. (1999)
we have $\tau_N \leq 7.5\times10^5$ years
for the North lobe, which accounts for most of the total radio luminosity.

A second way to determine the age of 3C99 is a dynamical approach via the
following relation:
\begin{equation}
<\tau> = \frac{D}{c <\beta> sin\theta}
\end{equation}
and for example for $<\beta_1>=0.05$,  $<\beta_2>=0.3$ and $\theta=30^\circ$ 
dynamical ages of $ 7 \times10^5$ and $4\times10^5$ for each 
lobe respectively, are derived. The two ages differ due to the time delay
in the propagation of the radiation.
\section{Conclusions}
The present observations confirm the asymmetry of 3C99. A one-sided long,
thin jet was detected with MERLIN. The 5-GHz MERLIN observations show that the 
source is polarized in the hot spot region and confirm the depolarization
previously measured, and a PA of the electric vector in agreement 
to previous results. The hot spot region 
contains three compact features detected by the 1.6\,GHz VLBA observations, 
the jet was resolved out and the central
area shows a double structure in both 1.6\,GHz and 5\,GHz VLBA images. 
The observations did not allow us to detected polarized
emission from any components in 3C\,99 above the detection limit of the
VLBA observations (0.3 mJy/beam at 5\,GHz).
The comparison between the 1.6\,GHz and the 5\,GHz shows that the detected
source components do show a steep spectral index, making it difficult to
suggest which of them is the core. We suggest that the
two components detected in the central region of 3C\,99, are co-moving
jet components along the north--east side of the source.

The asymmetry in the radio structure of 3C\,99 can be explained with the
Scheuer-Baldwin model requiring a rather large difference in the
interstellar medium density on the two sides of the central component.
Such an assumption makes the model predictions consistent with the
observational parameters.
We have assumed that in the south--west direction,
where the hot-spot is less bright and more distant, the ambient gas density 
goes as $r^{-2}$, where $r$ is the radial distance from the source nucleus . 
On the opposite side, 
i.e. in north--east direction, we have assumed that the ambient gas density is
constant with $r$, with a stronger interaction with the advancing jet.
This should also explain the strong far infrared emission detected.
The interaction of the radio plasma with the interstellar
medium may trigger the star formation and the heating of dust will generate
strong infrared emission.
However, relativistic effects were also required to fit the parameters
derived from the observations. An angle
of $30^\circ-40^\circ$ between the source major axis and the observers
line of sight is needed.

The age of 3C\,99 has been derived using both the age of the emitting 
relativistic electrons with a dynamical approach. The source is found
younger than $10^6$ years.

A time scale of $\sim 10^5-10^6$ has been suggested recently by
Baker et al. (2002) for the ionization cone to emerge from their dusty
cocoons in quasars which is comparable to those of many CSS and,
of course, to that of 3C99. They show that the stronger C\,IV absorbers are
found in small steep-spectrum radio sources and argue that the absorption
column density changes with time. Since it is unlikely that CSS
sources are confined by dense clouds of gas and since the environmental 
density cannot account for the full range of radio source size in quasars,
they suggest that the age must be important. Models of dust grains in the 
vicinity of powerful radio sources requires special conditions for grain 
survival up to $\sim 10^7$ years. According to such interpretation, 3C99 
can be understood to be in a phase in which the jets are emerging from the 
clouds of dense gas. 

We note that, though 3C\,99 has been classified as a N-galaxy, 
an overview of the
spectroscopy available at several wavelength rises new puzzling
questions about its nature in the light of the unified models. Its high 
infrared luminosity, which is a factor of 20 greater than that of a 
typical radio galaxy, and its optical spectrum which shows high luminosity 
narrow lines, would classify 3C\,99 as FIR-bright AGN. The lack of broad 
band optical emission and the rather weak X-ray emission, on the other hand,
are against such interpretation. New observations are needed to
confirm the alternative suggestion for the star formation process as 
origin of the infrared emission.

\begin{acknowledgements}
The author whish to thank the referee for valuable comments on the manuscript.
\end{acknowledgements}


\begin{thebibliography}{ }

\bibitem[2002]{Baker}
 Baker J.C., Hunstead R.W, Athreya R.M. et al 2002 ApJ, 568, 592
\bibitem[1982]{Baldwin}
 Baldwin J.E., 1982, in {\it Extragalactic Radio Sources},
 Proceeding IAU Symposium No.\,97, eds. D.S. Heeschen \& C.M. Wade,
 Dordrecht, Reidel, p.\,21
\bibitem[1996]{Begelman}
 Begelman M.C. 1996, Cyg\,A: Study of a Radio Galaxy, eds. C. Carilli \&
 D. Harris, Cambridge University Press, p.\,209 
\bibitem[1991]{Carilli}
 Carilli C.L., Perley R.A., Dreher J.W. et al. 1991 ApJ, 383, 554
\bibitem[1996]{Crawford}
 Crawford C.S. \& Fabian A.C. 1996, MNRAS 282, 1483
\bibitem[1995]{Fanti}
 Fanti C., Fanti R., Dallacasa D. et al. 1995, A\&A 231,333
\bibitem[2000]{Fanti}
 Fanti C., Pozzi F., Fanti R. al. 2000, A\&A 358,499 
\bibitem[1994]{Hes}
 Hes R. 1994, PhD Thesis
\bibitem[1995]{Hes}
 Hes R., Barthel P.D. \& Hoekstra H. 1995, A\&A 303,8
\bibitem[1996]{Hes}
 Hes R., Barthel P.D. \& Fosbury R.A.E. 1996, A\&A 313,423
\bibitem[1997]{Kaiser}
Kaiser C.R. \& Alexander P. 1997, MNRAS 286,215
\bibitem[1990]{Mantovani}
 Mantovani F., Saikia D.J., Browne I.W.A. et al. 1990, MNRAS 245,427
\bibitem[1997]{Mantovani}
 Mantovani F., Junor W., Fanti R. et al. 1997, A\&AS 125,573
\bibitem[1999]{Murgia}
 Murgia M., Fanti C., Fanti R. et al. 1999, A\&A 345, 769
\bibitem[1995]{Saikia}
 Saikia D.J., Jeyakumar S., Wiita P.J. et al. 1995, MNRAS 276, 1215
\bibitem[1996]{Saikia}
 Saikia D.J., Jeyakumar S., Wiita P.J. et al. 1996, 
 {\it The Second Workshop on Gigahertz Peaked Spectrum and Compact Steep
  Spectrum Radio Sources}, Leiden, 30 September - 1,2 October 1996
  Eds. I.A.G. Snellen, R.T. Schilizzi, H.J.A. R\"ottgering and M.N. Bremer  
\bibitem[1974]{Scheuer}
 Scheuer P.A.G. 1974, MNRAS 166, 513
\bibitem[1995]{Shepherd}
 Shepherd, M.C., Pearson, T.J., Taylor, G.B. 1995, BAAS 26, 987
\bibitem[1985]{Spinrad}
 Spinrad H., Djorkovski S., Marr J. \& Aguilar L. 1985, PASP 97,932
\bibitem[2001]{Bemmel}
 van Bemmel I. \& Barthel 2001, A\&A 379,L21
%
\end{thebibliography}
\end{document}